\documentclass{aa}  

\usepackage{graphicx}
\usepackage{txfonts}
\begin{document}

   \title{The Near-Infrared Spectrograph (NIRSpec) on the \textit{James Webb }Space Telescope\thanks{The eMPT software package and its associated user guide are available for download from the ESA GitHub page: https://github.com/esdc-esac-esa-int}}

   \subtitle{V. Optimal algorithms for planning multi-object spectroscopic observations}

\author{ N. Bonaventura\inst{1,}\inst{2}
\and P. Jakobsen        \inst{1,}\inst{2}
\and P. Ferruit         \inst{3}
\and S. Arribas         \inst{4}
\and G. Giardino        \inst{5}
}

\institute{ 
Cosmic Dawn Center, University of Copenhagen, Denmark \and
Niels Bohr Institute, University of Copenhagen, Jagtvej 128, DK-2200 Copenhagen, Denmark \and
European Space Agency, European Space Astronomy Centre, Madrid, Spain \and
Centro de Astrobiologia (CSIC-INTA), Departamento de Astrofisica, Madrid, Spain \and
ATG Europe for the European Space Agency, European Space Research and Technology Centre, Noordwijk, The Netherlands 
}

\date{Received Soon 2022; accepted Later 2022}

\abstract {We present an overview of the capabilities and key algorithms employed in the so-called eMPT software suite developed for planning scientifically optimized, multi-object spectroscopic (MOS) observations with the Micro-Shutter Array (MSA) of the Near-Infrared Spectrograph (NIRSpec) instrument on board the \textit{James Webb} Space Telescope (JWST), the first multi-object spectrograph to operate in space. NIRSpec MOS mode is enabled by a programmable MSA, a regular grid of $\sim$250,000 individual apertures that projects to a static, semi-regular pattern of available slits on the sky and makes the planning and optimization of an MSA observation a rather complex task. As such, the eMPT package is offered to the NIRSpec user community as a supplement to the MSA Planning Tool (MPT) included in the STScI Astronomer's Proposal Tool (APT) to assist in the planning of NIRSpec MOS proposals requiring advanced functionality to meet ambitious science goals. The eMPT produces output that can readily be imported and incorporated into the user's observing program within the APT to generate a customized MPT MOS observation. Furthermore, its novel algorithms and modular approach make it highly flexible and customizable, providing users the option to finely control the workflow and even insert their own software modules to tune their MSA slit masks to the particular scientific objectives at hand.}

               
\maketitle

\titlerunning{NIRSpec on the James Webb Space Telescope V}
\authorrunning{N. Bonaventura et al.}

%

\section{Introduction}
\label{sec:intro}

The design of the Near-Infrared Spectrograph (NIRSpec) instrument on board the \textit{James Webb} Space Telescope (JWST) and its different observing modes are presented in \cite{jak22}. As described in further detail by \cite{ferr22}, a key capability of the instrument is the multi-object spectroscopic (MOS) capability afforded by its novel Micro-Shutter Array (MSA). The MSA carries a total of $4 \times 365 \times 171 = 249\,660$ individually addressable micro-shutters arranged in four quadrants spanning a combined 3\farcm6$\times$3\farcm4 field of view on the sky. The nontrivial task of matching the regular grid of MSA shutters to a given ensemble of candidate targets on the sky in an optimal manner requires the use of bespoke specialized software -- and indeed, comprises the NIRSpec-specific portions of the Astronomer's Proposal Tool (APT)\footnote{http://apt.stsci.edu} in the so-called MSA Planning Tool \citep[MPT]{kara14}.

However, during the development and planning of the European Space Agency (ESA) NIRSpec Science Team's Guaranteed Time Observer (GTO) program, a number of ideas arose for ways to further optimize the scientific potential of the MSA. This prompted the development of the so-called eMPT software suite, which began as a test bed for such alternative approaches that are not currently possible. As many of these ideas proved to be quite powerful, the eMPT has since evolved into an alternative to the MPT and is being offered to the JWST user community on an ``as is'' basis (Sect.~\ref{sec:sw}). The purpose of this paper is to outline the different approach taken by the eMPT in planning MSA observations, list the unique capabilities enabled by this approach, and document the machinery of some of the key algorithms developed for it. The intent is not to repeat here the details of the eMPT User Guide, but to highlight what the eMPT may have to offer NIRSpec users for their particular MSA programs.


\section{Architecture of the eMPT}

Given a user-provided catalog listing the celestial coordinates and relative scientific priorities of the objects to be observed with the NIRSpec MSA, the common task of the MPT and eMPT software is to identify the precise MSA sky pointing required, and shutters that need to be opened, to observe the optimal number of highest-priority spectra in a single exposure, without the spectra overlapping on the detector. For reasons explained in Sect.~\ref{sec:roll}, this optimization is performed at a predetermined MSA orientation, but ultimately at the official roll angle assigned by STScI. This roll angle, the JWST ``aperture position angle'', refers to the position angle of the ``science y-axis'' of, in this case, the NIRSpec MSA on-sky field of view (FOV), measured eastward from north.

\subsection{Modular approach}

\begin{table*}
\caption{Basic modules of the eMPT.} 
\label{tab:modules} 
\centering
\begin{tabular}{ll}
\hline
\hline
\noalign{\smallskip}
Module& Function \\
\noalign{\smallskip}
\hline
\noalign{\smallskip}
{\tt ipa}& Find the MSA pointings within a user-defined search box that capture the maximum number of highest priority targets\\
{\tt k\_make}& For each optimal pointing identify all targets of all priority classes that fall within viable slitlets on the MSA (the raw {\it k\_list})
\\
{\tt k\_clean}& Purge the candidate list at each pointing of targets contaminated by nearby catalog entries\\
{\tt m\_make}&  Identify the optimal subset of targets whose spectra fit on the detector without overlap for each pointing (the filtered {\it m\_list})\\
{\tt m\_sort}& Rank the pointings and matching target sets according to a user-defined figure of merit\\
{\tt m\_pick}& Generate the information needed to import the highest ranking MSA mask and pointing combination into the APT \\
{\tt m\_check}& Create further diagnostic files and plots for the chosen solution\\
\noalign{\smallskip}
\hline
\end{tabular}
\end{table*}

The eMPT was built as a series of modules designed to be run in strict sequence, such that each module performs a specific task on the output of the previous module. All modules were written in standard \textit{gcc gfortran}, in order to keep the execution time associated with the most computationally intensive modules manageable. The models and primary functions of the baseline eMPT module suite are listed in Table~\ref{tab:modules}. The eMPT is controlled by a user-edited configuration file that simultaneously serves to set the parameters of each step, and provides a record of the total workflow. All intermediate results generated by each module are output as plain ASCII files that can be accessed and optionally modified by users either manually or with their own software (Sect.~\ref{sec:custom}). The eMPT modular work flow can also easily be captured in simple shell scripts (Sect.~\ref{sec:script}), which, together with the high speed of execution, encourages experimentation and active exploration of a variety of ``what if?'' scenarios.

\subsection{NIRSpec instrument model}

To carry out its functions, the eMPT requires a number of prerequisites, many of which are common with the MPT; one key element, which forms the backbone of both systems, is the NIRSpec instrument model \citep{dorn16,birk16,giov16,alve22,nora22}. The in-flight, updated version of the NIRSpec instrument model is capable of tracing light originating from a point on the sky, through the telescope and instrument, and finally onto the detector in any of the NIRSpec observing modes, to the accuracy of a fraction of a pixel \citep{nora22}. Particularly relevant to the eMPT is the polynomial fit employed by the model to capture the significant optical field distortion that occurs in the telescope and NIRSpec foreoptics between the sky and the MSA, along with the detailed metrology model of the MSA itself that specifies where precisely all its shutters lie in the NIRSpec slit plane. The NIRSpec model has been maintained and kept up-to-date by the ESA NIRSpec team until the end of JWST commissioning; future updates will be put in place by the STScI NIRSpec group. The eMPT and MPT use the same set of model reference files to capture its most recent version.

\subsection{MSA shutter operability and viable slitlet maps}
\label{sec:msamap}

Another important component of the MPT and eMPT is the MSA shutter operability map, which keeps a tally of which shutters are fully functional, failed closed, and failed open at a given point in time \citep{rawl22}. The eMPT is presently hard-wired to assume that each target to be observed with the MSA is assigned a three-shutter-tall slitlet, and that observations are nodded between sub-exposures by shifting the MSA pointing up and down by one shutter. Central to the eMPT approach is the so-called {viable slitlet map}, a $4\times365\times171$ look-up table that identifies a given shutter $(k,i,j)$ as capable of serving as the central shutter of a fully functioning, three-shutter-tall slitlet. This viable slitlet map is initialized at the start of each eMPT run and is then continuously updated as part of the workflow, as targets and their spectra are added to the MSA and detector. The initialization process depends on the disperser being used, and involves identifying and excluding from use any viable slitlets whose spectra collide with those of failed open shutters on the MSA. In the case of NIRSpec's low-resolution (R$\sim$100) prism disperser (PRISM), the initialization process also includes precluding from use those slitlets that result in incomplete spectra (Sect.~\ref{sec:prism_overlap}).

\subsection{Tangential coordinates}

When projecting targets on the sky through the JWST telescope onto the NIRSpec MSA, the eMPT makes use of the so-called {tangential} (or standard) angular approximation throughout \citep{vdk67}. In this approximation, a point on the sky at spherical coordinates ($\alpha$,$\delta$) is, for a given MSA pointing ($\alpha_p$,$\delta_p$), represented by the relative Cartesian angular coordinates ($x_\alpha$,$y_\delta$), where
\begin{equation}
x_\alpha = \frac
{\cos\delta \sin(\alpha-\alpha_p)}
{\sin\delta \sin\delta_p + \cos\delta \cos\delta_p \cos(\alpha-\alpha_p)} 
\label{eq:tcoordx}
\end{equation}
\begin{equation}
y_\delta = \frac 
{\sin\delta \cos\delta_p - \cos\delta \sin(\alpha-\alpha_p)}
{ \sin\delta \sin\delta_p + \cos\delta \cos\delta_p \cos(\alpha-\alpha_p)}   
\label{eq:tcoordy}
.\end{equation}
Here $x_\alpha$ and $y_\delta$ respectively denote the west-to-east and south-to-north angular distances from the point ($\alpha_p$,$\delta_p$) to ($\alpha$,$\delta$). This tangential approximation is accurate to a fraction of a milliarcsecond (mas) over the  3\farcm6$\times$3\farcm4 field of view of the MSA, well below the NIRSpec instrument model accuracy. Consequently, $x_\alpha$ and $y_\delta$ can be rotated to the orientation of the telescope and treated as customary planar Cartesian coordinates. 
When the need arises to convert tangential coordinates into proper spherical coordinates, the inverse expressions employed are 
\begin{equation}
\alpha = \alpha_p +\arctan(\frac
{x_\alpha}
{\cos\delta_p+y_\delta\sin\delta_p})
\end{equation}
\begin{equation}
\delta = \arcsin(\frac
{\sin\delta_p + y_\delta \cos\delta_p }
{\sqrt{1 + x_\alpha^2 + y_\delta^2}})
.\end{equation}

\subsection{Scientific prioritization of targets}

Compared to the MPT, the eMPT employs a rather coarse prioritization of the target candidates, grouping the catalog entries into only a handful of scientific priority classes (typically fewer than 20); there are several reasons for this choice. First and foremost, it reflects the scientific objectives of most real-world MSA programs, that is to say, one is faced with a small number of extremely high-priority targets for which it is vital to obtain the greatest possible number of spectra, after which the MSA should ideally be sequentially filled with the greatest possible number of more abundant, but decreasingly important, targets. In fact, this means that the algorithm employed in the eMPT to pack the largest number of non-overlapping spectra on the NIRSpec detector array (Sect.~\ref{sec:arribas}) becomes more efficient with more (lower priority) targets to consider. 

The targets designated as Priority Class 1 (PC1) in the eMPT are treated in a very different manner from the remaining targets to exploit one of the most powerful features of the eMPT, its so-called initial pointing algorithm (IPA). The IPA module takes as input the user-supplied catalog containing a number of labeled PC1 targets, a nominal MSA pointing position on the sky ($\alpha_o,\delta_o$), together with a specification of the size of the maximum allowed deviations from this nominal pointing. From this input the IPA produces, for the assigned fixed roll angle $\phi_c$, a list of detailed MSA pointings ($\alpha_p,\delta_p,\phi_p$) within the specified search box that provide the largest possible number of non-overlapping spectra of PC1 targets. The way in which the IPA module achieves this feat is described in Sect.~\ref{sec:ipa}. It is important to appreciate that once the IPA module has produced its list of PC1-maximizing pointings, these are the {only} pointings subsequently considered by the following, downstream eMPT modules in Table~\ref{tab:modules}.\ They together endeavor to fill the MSA at each pointing identified by the IPA with an optimal set of lower priority targets, and, in the end, identify which of these target sets is the most scientifically preferred overall. 

\section{Avoiding spectral overlap}
\label{sec:overlap}

Another prerequisite common to the eMPT and MPT is a means of quickly assessing whether the spectra arising from two shutters on the MSA overlap on the NIRSpec detector array, both vertically in the spatial direction and horizontally along the dispersion direction. 
As explained in \cite{jak22}, NIRSpec grating spectra are straddled horizontally by zero- and second-order spectra at their blue and red ends and therefore effectively span the entire width of the detector array. Horizontal spectral separation is therefore only of concern for the shorter, lower-resolution PRISM spectra, of which as many as four can fit across the detector.

In an ideal instrument, the separation between two spectra from different shutters would simply reflect the horizontal and vertical differences, $\Delta i$ and $\Delta j$, between the two shutter locations on the MSA. However, optical distortion in the NIRSpec collimator optics and, to a lesser extent, small misalignments between the four MSA shutter quadrants and residual spatial clocking differences in the mounting of the NIRSpec dispersers, significantly complicate the situation.

\subsection{Vertical overlap}

\begin{figure}
 \centerline{ \resizebox{6.5truecm}{!}{\includegraphics{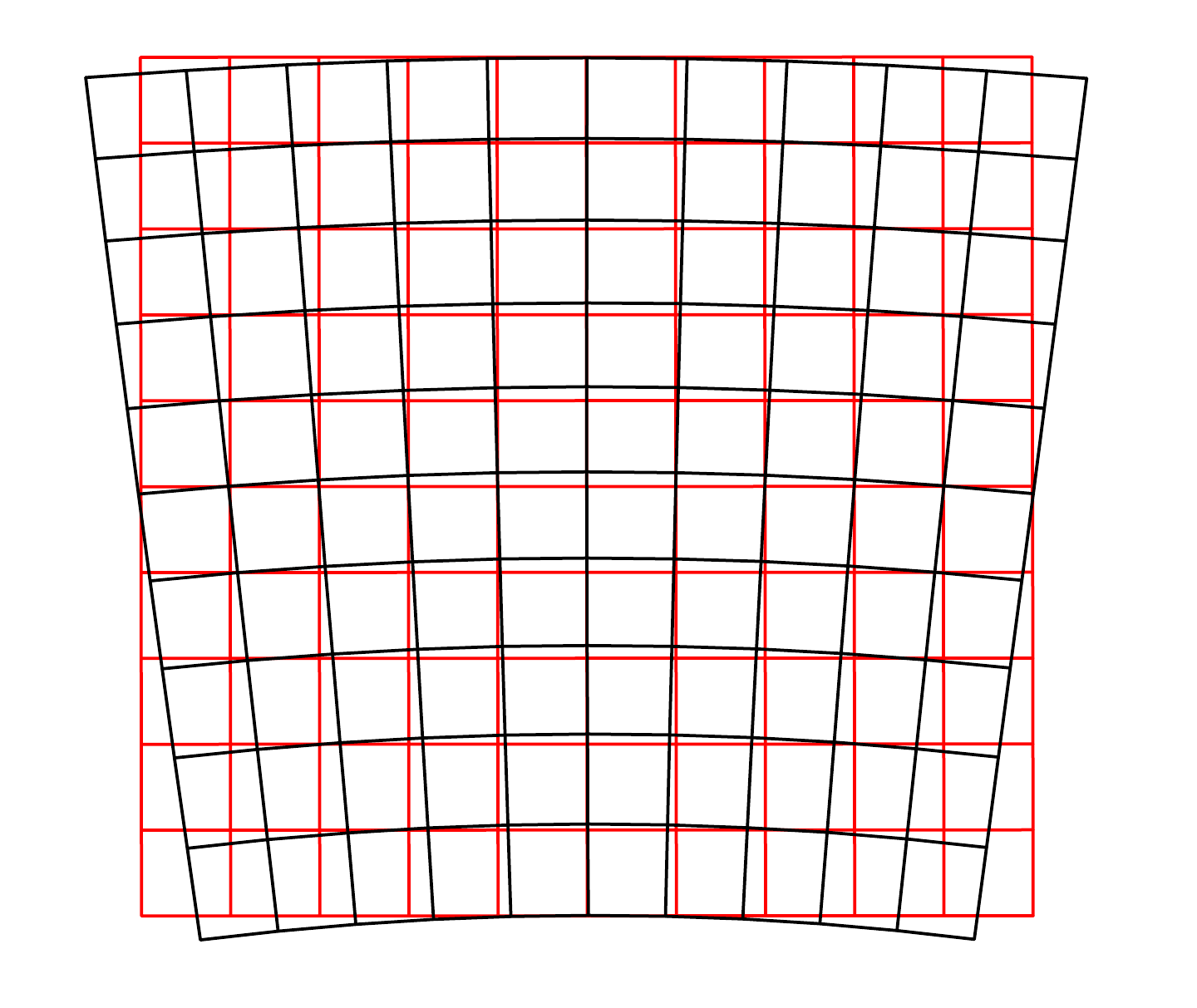}}}
  \caption{Map of the angular field distortion displayed by the NIRSpec collimator optics. The best-fit linear model is shown in red. The actual distortion, amplified by a factor of 25 for clarity, is plotted in black.}
  \label{fig:col_dist}
\end{figure}

The optical field distortion in the angular output of the NIRSpec collimator is shown in exaggerated form in Fig.~\ref{fig:col_dist}. In this figure, the upward curvature of the horizontal lines, corresponding to input along a fixed MSA shutter row, results in spectra from MSA shutters located toward the middle of the MSA to be imaged $\simeq\!4\!-\!6$ pixel rows higher on the detector relative to spectra from shutters near the two edges. As a consequence, the vertical separation between the spectra from any two shutters cannot be simply gauged by the separation in shutter rows, $\Delta j$. Following a suggestion made by the STScI NIRSpec team, this complication is handled in both the eMPT and MPT by mapping the distortion onto a single scalar $S(k,i,j)$ assigned to each shutter of the MSA, such that the vertical spectral separation measured in equivalent shutter rows can instead be accurately assessed simply by differencing the two so-called {shutter values}.  

The dominant contribution to $S(k,i,j)$ stemming from the collimator distortion can then be calculated as
\begin{equation}
S_{\!C}(k,i,j) = j + \left({f_y^{COL} \over p_y}\right) \Delta\theta_y(k,i,j)
\label{eq:coll_sv}
,\end{equation}
where $\Delta\theta_y(k,i,j)$ is the deviation in the vertical exit angle of the beam originating from shutter $(k,i,j)$ compared to an ideal linear optical alignment; $f_y^{COL}=660$~mm is the mean focal length of the collimator in the spatial direction; and $p_y=0.204$~mm is the shutter pitch in the spatial direction. 

A second, smaller contribution to $S(k,i,j)$ comes from small residual misalignments in the MSA, which, in the worst case, amount to approximately half a shutter pitch in the spatial direction,
\begin{equation}
S_{\!M}(k,i,j) =  \left({1\over p_y}\right) \Delta y_m(k,i,j)
\label{eq:msa_sv}
,\end{equation}
where $\Delta y_m(k,i,j)$ is the spatial displacement of shutter ($k,i,j)$ with respect to the perfectly aligned MSA.

\begin{figure}
  \centerline{\resizebox{8.5truecm}{!}{\includegraphics{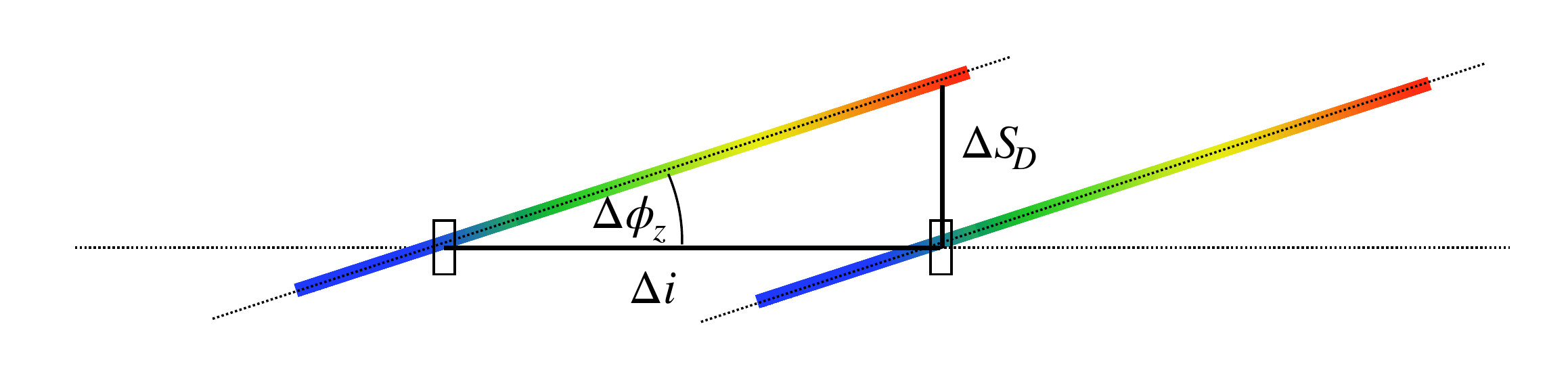}}}
  \caption{Schematic illustration of the vertical separation, $\Delta S_{\!D}$, between two spectra from two shutters separated horizontally by $\Delta i$ when the disperser is mis-clocked by the angle $\Delta\phi_z$}
  \label{fig:misclock}
\end{figure}

Whereas the two above contributions to $S(k,i,j)$ are geometrical in nature, the third contributor depends on the disperser being deployed. As illustrated in Fig.~\ref{fig:misclock}, a grating or prism is rotated in its mount by an angle, $\Delta \phi_z$, which leads to its dispersed spectra being tilted by the same angle with respect to the detector pixel rows. In this case, the vertical separation between two spectra from shutters located in the same MSA row grows in proportion to the horizontal separation between the shutters. The contribution to $S(k,i,j)$ from this effect can be written as
\begin{equation}
S_{\!D}(k,i,j) = \left( p_x\over p_y\right) \tan(\Delta\phi_z) \thinspace i^\star(k,i,j) 
\label{eq:clock_sv}
,\end{equation}
where $p_x=0.105$~mm is the horizontal shutter pitch, and $i^\star(k,i,j)$ accounts for cases where the shutters lie in adjacent quadrants,
\begin{equation}
i^\star(k,i,j) = \begin{cases}
i                    &\text{ if  $k=1$ or $2$}\\
i+\Delta_i^{gap}+365 &\text{ if  $k=3$ or $4$}
\end{cases}
\label{eq:i_star_sv}
,\end{equation}
where $\Delta_i^{gap}=85.98$ is the equivalent number of horizontal shutters spanned by the vertical gap between shutter quadrants. Values of $\Delta \phi_z$ for the seven NIRSpec dispersers are listed in Table~\ref{tab:clocking}. It is seen that the six gratings are  well-aligned to within approximately an arcminute, except for the high-resolution G235H grating, which stands out at $\Delta\phi_z=-350$\arcsec. The prism is mis-clocked at an even greater angle, but since its spectra are much shorter, this is less noticeable. Nonetheless, the differences in $\Delta\phi_z$ in Table~\ref{tab:clocking} are still sufficient to require that each disperser have its own separate shutter value table, formed from the sum of the contributions of Eqs. (\ref{eq:coll_sv}) and (\ref{eq:msa_sv}) and the disperser-dependent Eq. (\ref{eq:clock_sv}):
\begin{equation}
S(k,i,j)=S_{\!C}(k,i,j)+S_{\!M}(k,i,j)+S_{\!D}(k,i,j)
\label{eq:sum_sv}
.\end{equation}
The eMPT therefore employs seven separate shutter value tables calculated for each disperser per the above recipe. In the interest of employing fewer such tables, the MPT trades off accuracy by sharing tables between the dispersers in Table~\ref{tab:clocking}, whose mis-clocking angles are not too dissimilar. {In practice, the vertical spectral collision algorithms employed by the MPT and eMPT are identical.}

\begin{table}
\caption{Disperser clocking angles.} 
\label{tab:clocking} 
\centering
\begin{tabular}{cc}
\hline
\hline
\noalign{\smallskip}
Disperser& $\Delta \phi_z$ [arcsec] \\
\noalign{\smallskip}
\hline
\noalign{\smallskip}
G140M& +62\\
G235M& +82\\
G395M& +6\\
G140H& -23\\
G235H& -350\\
G395H& +30\\
PRISM& +1060\\
\noalign{\smallskip}
\hline
\end{tabular}
\end{table}

\subsection{Horizontal overlap of PRISM spectra}
\label{sec:prism_overlap}

\begin{figure}
  \resizebox{\hsize}{!}{\includegraphics{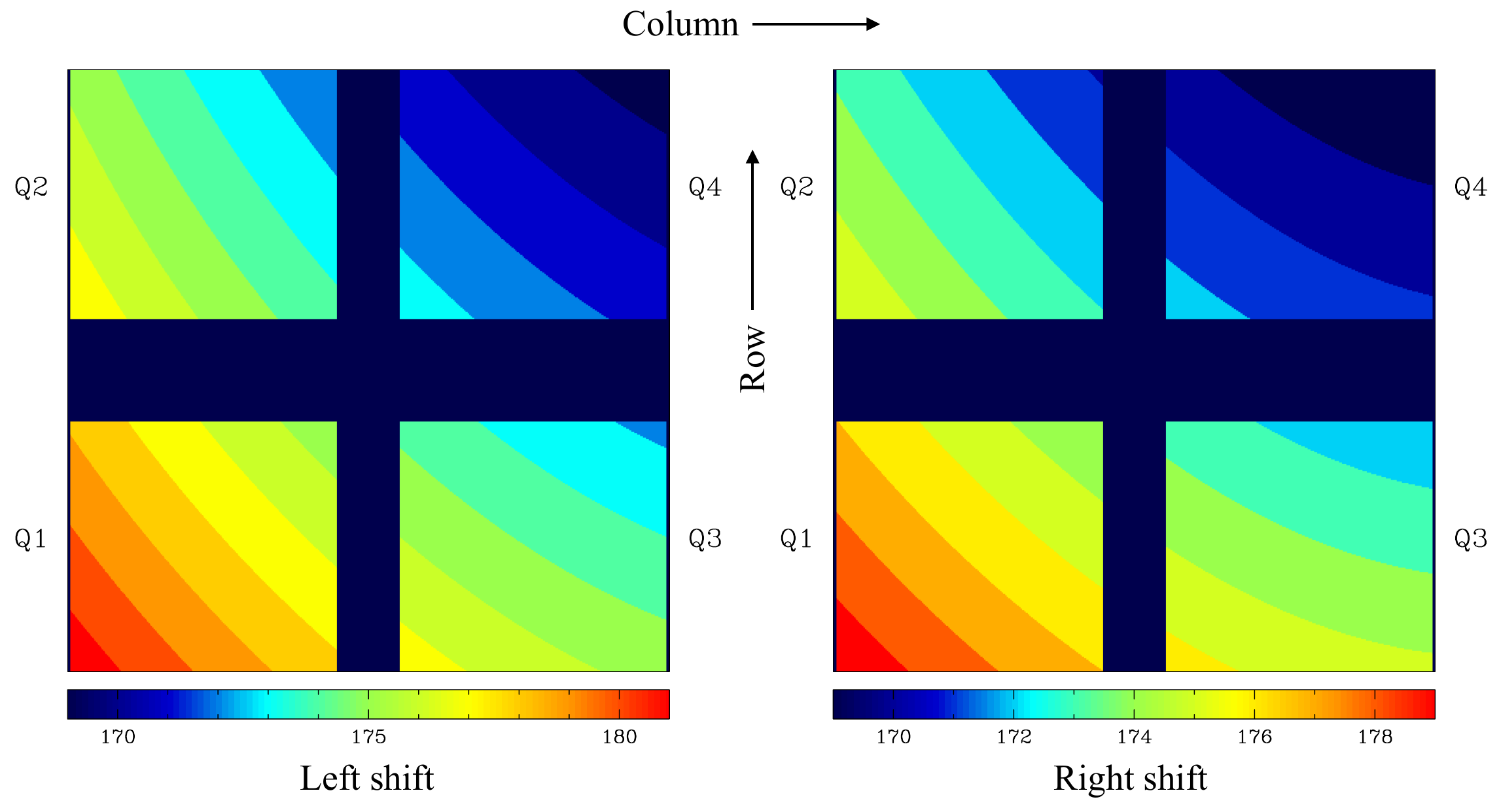}}
  \caption{Maps of how many shutters to the left and right along an MSA row are required for PRISM spectra to avoid overlapping horizontally.}
  \label{fig:prism_sep}
\end{figure}

Spectra obtained with the NIRSpec low-resolution prism disperser do not exhibit spectral orders; these spectra extend from $\lambda\!\simeq\!0.5~\mu$m up to $\lambda\!\simeq\!5.5~\mu$m and have a length that, depending on the shutter employed, varies between 172 and 184 pixels across the face of the MSA. Consequently, up to four prism MSA spectra can fit across the NIRSpec detector arrays without overlapping or being truncated by the gap between the two arrays (Fig.~\ref{fig:prism_spectra}). In order to quickly asses whether the PRISM spectra from two shutters $(k,i,j)$ and $(k',i',j')$ overlap horizontally, the eMPT carries two precalculated $4\times365\times171$ integer look-up tables that specify how far away $i'$ must be located respectively to the left and right of shutter $(k,i,j)$ for the red and blue ends of the two spectra to clear each other (Fig.~\ref{fig:prism_sep}). By taking into account the slight asymmetry in the PRISM spectra locations in this manner, the eMPT ensures that the greatest possible number of such spectra can be placed on the NIRSpec detector arrays. In contrast, the MPT only employs a single minimum required horizontal shutter column separation threshold, which of necessity is the maximum value in the two maps shown in Fig.~\ref{fig:prism_sep}.  {In practice, however, this difference is insignificant except at the highest target densities (Fig.~\ref{fig:prism_spectra}), where, with all other factors being equal, it enables the eMPT to pack $\sim$1\% more non-overlapping PRISM spectra on the detector compared to the MPT.}

\section{The matrix ({Arribas}) algorithm}
\label{sec:arribas}

A unique core element of the eMPT is the optimal manner in which it decides which subset of a given ensemble of targets located within viable slitlets on the MSA produces the largest possible number of non-overlapping spectra on the NIRSpec detector array. This critical down-selection step is carried out throughout the eMPT by employing the so-called matrix algorithm (historically known as the {Arribas algorithm}). When supplied with a list of viable slitlets containing targets, the algorithm uses the methods outlined in the previous Sect.~\ref{sec:overlap} to quickly map which target spectra collide with others on the detector. The algorithm then identifies which candidate target clashes with the most competing other spectra and repeatedly eliminates the ``worst offender'' from the list until no spectral clashes remain. The final surviving set of targets is then the optimal one providing the largest possible number of spectra that can be drawn from the input set.

The matrix algorithm is employed in both the {\tt ipa} module, to eliminate spectral collision among the PC1 targets; and also in the {\tt m\_make} module, where the list of in-slit targets for each candidate pointing identified and purged of contamination by the {\tt k\_clean} module (Sect.~\ref{sec:contam}), are placed on the MSA, in the order of increasing priority class. This approach ensures that the final set of targets to be observed is the optimal one, providing the largest possible number of spectra of the highest-priority PC1 targets, followed by the largest possible number of spectra of Priority Class 2 spectra, conditional on the PC1 targets already being in place, in a successive fashion down through the lower-priority (higher-numbered) priority classes.

An interesting aspect of the matrix algorithm is how it handles ties between targets, that is, when two or more candidate targets conflict with the same number of competing targets. Arriving at the absolute optimal solution would in such cases require that all branches springing from all ties be pursued to completion, which is computationally demanding. Moreover, experiments in which the tied targets flagged for elimination were selected at random, do not suggest that there is much to be gained by mapping out such branches. In any event, since it is vital for testing purposes that the matrix algorithm not be stochastic, but repeatable for a given input, the eMPT employs the approach of always selecting for elimination the tied target that appears the furthest down in the user-provided input catalog. Therefore, among the otherwise equal targets assigned a given priority class, there is an implicit secondary ranking of their importance hiding in the order in which the targets are listed. With this additional constraint, the matrix algorithm is guaranteed to arrive at the best possible solution. 

In contrast to the eMPT's grouped approach, the MPT always works in a ``first come, first served'' manner by placing spectra on the detector one after the other, in strict sequence, working its way down the input catalog. {When there are few targets available and little or no spectral overlap, all targets falling within viable slitlets will make it onto the detector in either approach. However, as the number of targets that need to be accommodated is increased, and the likelihood of spectral overlap rises, the difference becomes more and more significant. In the limit of a very high target density, the matrix algorithm allows up to $\sim$17\% more spectra to be packed onto the detector array without overlap, compared to a single pass of the MPT approach. } However, the MPT does offer the computationally heavy option of randomly shuffling the order of the input catalog a number of times and selecting the best outcome amongst the trials. The matrix algorithm can therefore be thought of as the direct path to the result of having shuffled the input catalog a large number of times; this claim is also borne out quantitatively through direct testing. An illustrative example of what the matrix algorithm is capable of achieving is shown in Fig.~\ref{fig:prism_spectra}

\section{The initial pointing algorithm}
\label{sec:ipa}

Another powerful and unique feature of the eMPT is its IPA, designed to identify the optimal set of MSA pointings on the sky that provide the highest possible simultaneous coverage of the highest-priority PC1 targets in the input catalog, within a user-specified search box and for the assigned roll angle. The IPA accomplishes this by first employing an ``all-seeing'' formalism that is computationally manifested as a simultaneous ``mapping and stacking'' of the full projected, viable MSA shutter area accessible to all PC1 target positions within their respective search boxes, without relying on conventionally ``blind’' grid-searching or pattern-matching methods.  In the theoretical limit of infinitely precise telescope pointing accuracy and optics, this novel algorithm for optimizing MOS configurations locates the universal set of pointings that maximize the number of PC1 targets in a single observation. In reality, this algorithm must be implemented digitally, with a chosen digital sampling size that accounts for the telescope pointing accuracy and balances computational speed with desired results\footnote{Stress tests of the software for a variety of common use cases have thus far shown it to be a very rare occurrence that the currently adopted IPA digital sampling size is too coarse to locate the most optimal pointing solution. Experience has shown that sampling the digital shift map at one seventh of the width of the Acceptance Zone in the dispersion direction and one thirteenth of the height of the Acceptance Zone in the spatial direction is adequate to capture the optimal pointings. {This parameter is currently hardwired in the eMPT code but is easy to modify upon the individual user's request.}} It must also be fine-tuned to address the numerous complications arising from the specific observational setup at hand; in this case, the possibility of spectra overlapping on the NIRSpec detector array (Sect. 3); the differential optical field distortion occurring in the projection between the sky and MSA; and the slight change in MSA roll orientation that the APT imposes on each pointing shift (Sect. 7.1). Therefore, this digitized search is followed by a more precise, individual examination and fine-tuning of each of the possibly optimal pointings uncovered in the first step. The resulting two-pronged approach is both highly robust and very quick to perform.

\begin{figure}
  \resizebox{\hsize}{!}{\includegraphics{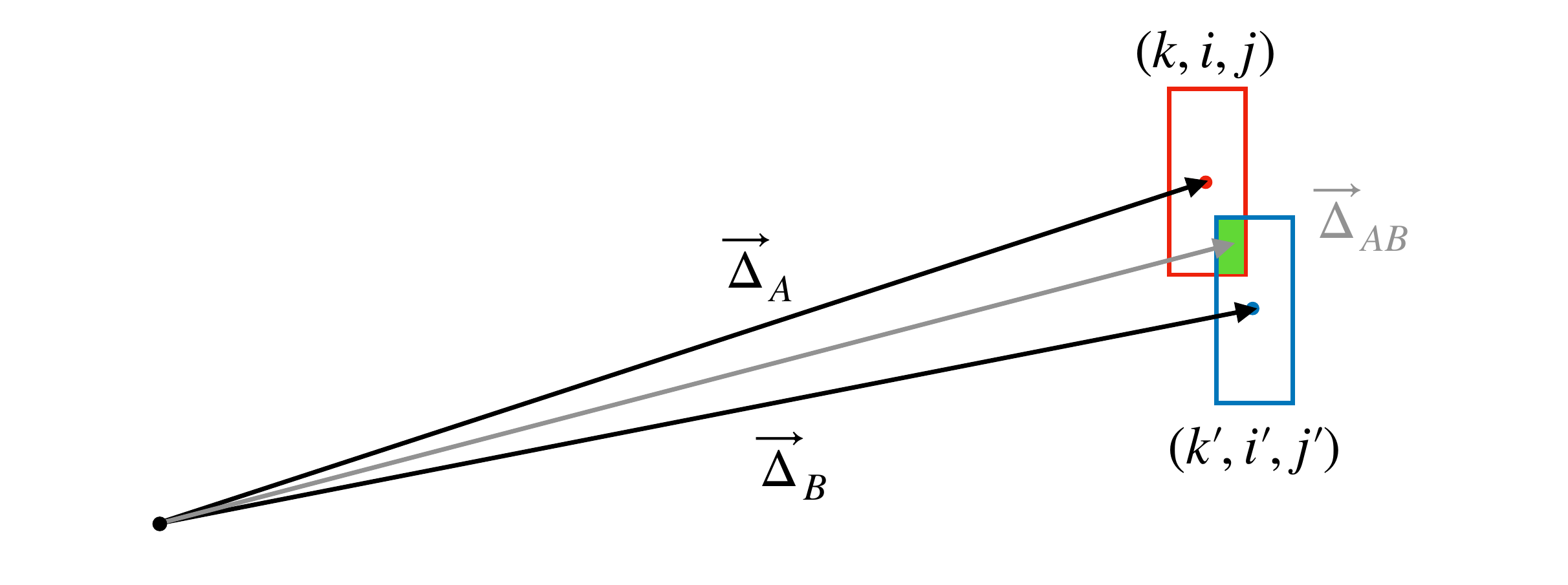}}
  \caption{Schematic illustrating the logic behind the digital shift-vector map employed in the IPA. Each rectangle shown represents the acceptance zone of a viable shutter (identified by its MSA coordinates) in shift-vector space, accessible to the associated target (red for target A and blue for target B) within a given search box. The (green) area of overlap of these acceptance zone areas in shift-vector space contains the shift vector value required from the nominal pointing position to simultaneously observe these two targets in a single MOS observation.}
  \label{fig:ipa_vector}
\end{figure}

The logic behind the first digital step of the IPA is illustrated in Fig.~\ref{fig:ipa_vector}. A target $A$ is located somewhere within the MSA FOV and a viable MSA shutter $(k,i,j)$ is accessible within the search box centered on the target; $\vec{\Delta}_A$ denotes the shift in pitch and yaw of the MSA pointing relative to the nominal pointing required to place target $A$ in the exact center of shutter $(k,i,j)$. For target $A$ to be observable in shutter $(k,i,j)$, the pointing shift need not be exactly $\vec{\Delta}_A$, but equal to any of the numerous discretized pointing shifts that place it within the shutter's acceptance zone.\footnote{The number of different shift vectors available to a given target to place it within the open area of a given shutter, grows with a decreasing digital sampling size of the shift vector space -- but at the cost of computational speed. The currently implemented size in the IPA is sufficiently large to allow for a short code running time, but small-enough to achieve the optimal pointing solution(s) and conveniently, clusters of closely spaced pointings to serve as dithered pointings.}  Considering now a second target $B$ located elsewhere on the MSA, requiring a shift $\vec{\Delta}_{B}$ to place it at the center of the viable shutter $(k',i',j')$: as is evident from Fig.~\ref{fig:ipa_vector}, if the open areas of the two shutters $(k,i,j)$ and $(k',i',j')$ happen to overlap in this shift-vector space, then an MSA shift by any compromise vector $\vec{\Delta}_{AB}$ ending in the green overlap zone will result in targets $A$ and $B$ being observable simultaneously, placing them respectively in the lower-right and upper-left corners of the two shutters $(k,i,j)$ and $(k',i',j')$. The first step of the IPA performs such a comparison between all priority class 1 targets in the input catalog by laying down an area corresponding to the acceptance zone of all possible target and viable shutter pairings in a digital shift map spanning the search area. The numerical peaks in this digital map then signal the shifted pointings that allow the greatest number of priority class 1 targets that can be observed simultaneously in a single MSA exposure.

However, due to the numerous NIRSpec observational challenges listed above, a local peak in the digital shift map should be considered as a necessary, but not sufficient, indicator of a true optimal-coverage pointing providing maximal simultaneous target coverage. The highest maxima detected in the idealized digital shift map are therefore in the IPA followed up individually in the second, more accurate fine-tuning step that takes into account these complications. For each peak in the digital map, the second step of the algorithm essentially starts from scratch, projecting the priority class 1 targets to the MSA using the initial MSA pointing $(\alpha_p,\delta_p)$ implied by the location of the peak in the digital shift map (and the MSA roll angle $\phi_p$ given by Eq.~\ref{eq:eq_of_e} below), and performing all the steps carried out by the full eMPT chain, including employing the matrix algorithm (Sect.~\ref{sec:arribas}) to eliminate spectral overlap and identify which targets are actually covered by the pointing. This process is then repeated iteratively until a self-consistent solution in which the final values of $(\alpha_p,\delta_p)$ have the outer envelope of the superposition of the intra-shutter locations of the covered targets perfectly centered within the imaginary common shutter. The surviving highest coverage pointings are then purged of duplicate pointings and, if necessary, sorted into groups according to the different lists of priority class 1 targets that they each cover. This last step is intended for cases where the optimal choice of two or three overlapping ``dithered'' pointings are being sought (Sect.~\ref{sec:mult}). Drawing overlapping pointings from the same group assures that the same priority class 1 targets are covered at all pointings.  

The MPT also has the capability to search for optimal pointings, but does so by carrying out a computationally intensive search over a grid of finite points spanning the search area. However, as is clear from Fig.~\ref{fig:ipa_vector} since pointings capturing a large number of targets simultaneously may extend over a very small region in search space, the blind MPT raster search generally needs to be performed with a very small step size to be assured of locating the optimal pointings that the IPA algorithm is designed to instantly isolate. 

\section{Other capabilities of the eMPT}
\label{sec:capa}
\medskip

\subsection{Allowing user customization}
\label{sec:custom}

The modular approach of the eMPT is designed to allow users to insert their own customized processing modules into the workflow. A prime use of this capability is to process the list of all target candidates that project to viable shutters at a given pointing (output by the {\tt k\_make} module) with the aim of flagging and excluding from further consideration those with known redshift or radial velocity whose spectra would be lacking coverage of certain emission or absorption lines of interest (due to the relevant portions of those spectra falling outside the active areas of the two NIRSpec detector arrays for the shutters in which they happen to lie). Precalculated wavelength coverage look-up tables for all seven NIRSpec dispersers can be provided for this purpose.

\subsection{Optimizing multiple exposures simultaneously}
\label{sec:mult}

The eMPT also has the ability to optimize two or three partially overlapping MSA pointings simultaneously; this is done by examining all suitably separated pairs or triples of pointings drawn from the list of PC1-optimized pointings produced by the {\tt ipa} module (Sect.~\ref{sec:ipa}). In the case of three dithered pointings, the common list of targets covered by the triplet are sorted for each priority class into targets that are a) common between all three pointings, b) targets that are observed in only two of the pointings, c) and those that are only observed in a single pointing of the triplet. Maximum commonality is then achieved by using the matrix algorithm (Sect.~\ref{sec:arribas}) to, in turn, place the thrice-covered targets on the MSA in order of increasing priority class, followed by the twice-covered targets, and lastly, the singly covered targets. Moreover, since the target placement sequence, and the weights that are assigned to priority class to evaluate the figure of merit of the target set covered by each triple, are fully under user control, the {opposite} goal of covering as many {different} targets as possible in the three pointings (or any intermediate case), is also possible. The opposite goal of covering as many different targets as possible in the three pointings (or any intermediate case), is also possible (even for PC1 targets by selecting pointings from different groupings found by the  {\tt ipa}).

In the MPT Planner, by comparison, the ``fixed'' dither option (distinct from nodding ``dithers'') is available for MOS observations. This option translates a planned slitlet configuration to a new location within the MSA field of view according to the user's specified offsets in the dispersion and/or cross-dispersion directions in shutter units, allowing for the selection of partially or fully observed primary sources across all dither pointings -- but not accounting for the differing positional shifts of different sources that will inevitably result from optical distortions between the dither pointings (thus the user-input dither offsets are considered ``average'' offsets between an initial and new pointing).

\subsection{Avoiding incomplete PRISM spectra}
\label{sec:prism_trunc}

The utilization of a dynamically updated viable slitlet map (Sect.~\ref{sec:msamap}) in the eMPT allows for the exclusion of MSA shutters that lead to PRISM spectra projecting onto the $\simeq156$~pixel-wide vertical gap between the two NIRSpec detector arrays, at the start of each eMPT run. The MPT does not censor such shutters and, as a result, typically 19\% of all PRISM spectra in MPT-designed MSA masks will be missing as much as half of their wavelength range. In the eMPT, truncated spectra are instead traded off against typically half as many complete PRISM spectra; the result of this censoring can readily be seen in Fig.~\ref{fig:prism_spectra}. This means that, for 100 PRISM spectra obtained by the MPT, $\sim$81 would be complete versus $\sim$91 obtained by the eMPT.

\subsection{Avoiding contaminated targets}
\label{sec:contam}

As its name suggests, the task of the so-called {\tt k\_clean} module between the {\tt k\_make} and {\tt m\_make} modules in Table~\ref{tab:modules} is to eliminate, early on in the down-selection process, any candidate targets whose spectra are likely to be contaminated by nearby objects{, according to the contamination rules detailed in Sect. 1.1 of the eMPT User guide}. The algorithm presently employed in the {\tt k\_clean} module is simplistic: for each target located within a viable slitlet, it projects its nodded slitlet to the sky and checks and flags if any of the five projected shutter areas contain any other catalog targets, regardless of their brightness. By specifying a set of thresholds in the eMPT configuration file, the user can control the overall level of such contamination that is deemed acceptable.  {In the present implementation, these threshold values merely stipulate the degree to which the centers of nearby contaminating sources, listed in the user-supplied input catalog, are allowed to be present in the open areas of the three shutters making up the slitlet at each of its three nods on the sky.}

This relatively crude ``point source'' approach to contamination elimination is ill-suited to handling extended objects whose signal may contaminate a given shutter even when its centroid position lies well outside of it; more sophisticated algorithms that involve projecting the slitlets onto an existing \textit{Hubble} Space Telescope image (or eventually a JWST image)  of the target field are under development. However, the present algorithm is expected to sufficiently eliminate the bulk of the major contamination sources in a first pass. In the MPT, contaminated targets are only flagged {after} an MSA configuration is constructed, in the list of observed targets covered by the pointing and MSA mask; they are not eliminated.

\subsection{Adding sky-only background spectra}

At any given pointing of the MSA on the sky, there will nearly always be a number of spare, unused slitlets available once the list of observable candidate targets in the input catalog has been exhausted. The eMPT is capable of exploiting this spare capacity by optionally filling the MSA with a near-optimal number of slitlets that are absent of targets. This allows the user to obtain as many as possible sky-only background spectra simultaneously with the target spectra, from which a ``master'' background spectrum for the exposure can be extracted and used in the reduction of the target spectra. Since faint-end MSA observations are limited by detector noise, this approach has the potential to notably decrease the amount of noise introduced during background subtraction.

\begin{figure}
  \resizebox{\hsize}{!}{\includegraphics{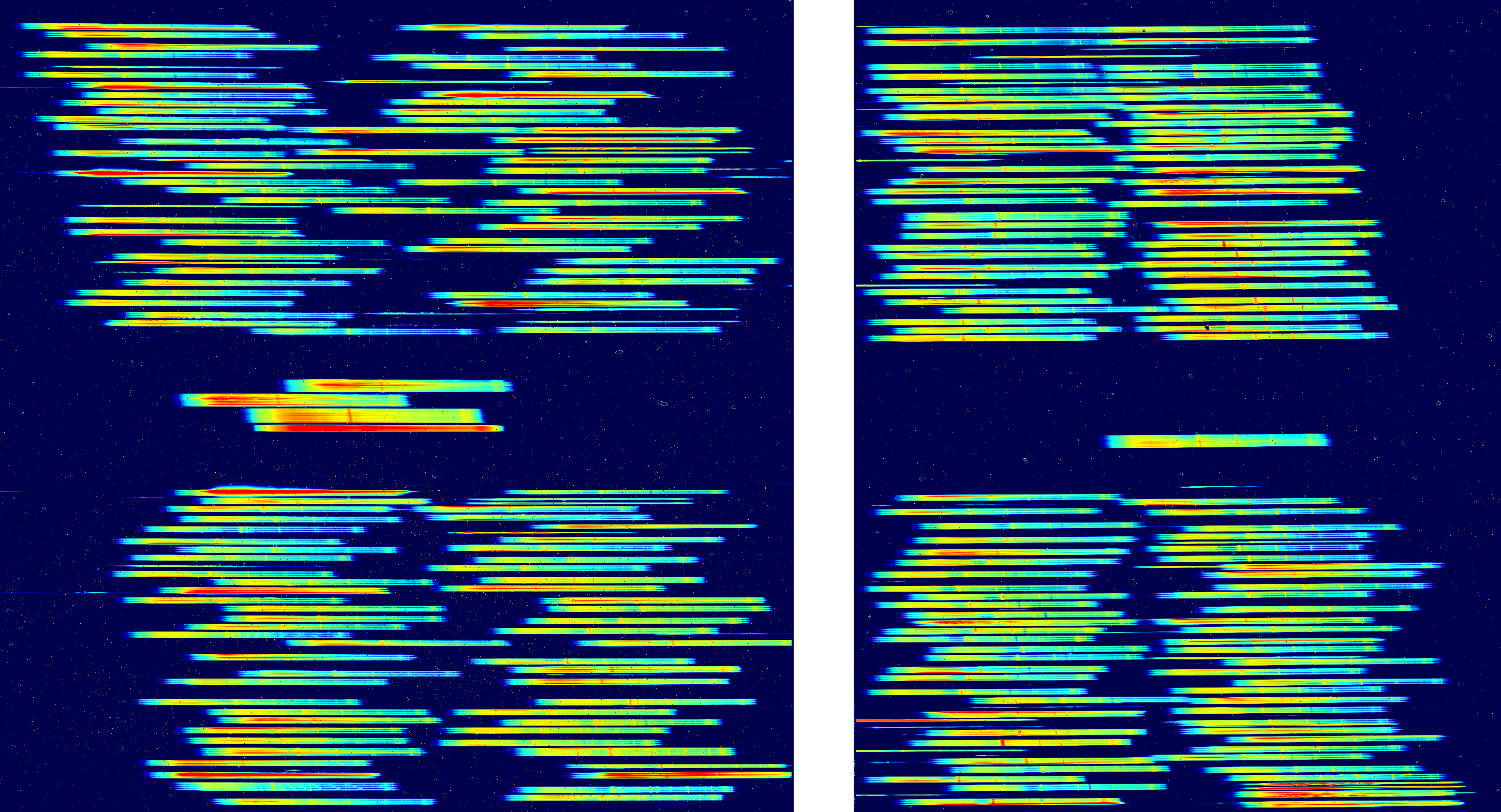}}
  \caption{Example of an MSA PRISM exposure obtained during NIRSpec commissioning in which the matrix algorithm, provided with the bare viable slitlet map as input, was used to produce an optimized MSA mask that yields a total of 235 non-overlapping and non-truncated MSA and 5 fixed-slit PRISM spectra of the diffuse interstellar medium emission during a blind pointing toward the field near the Galactic center.}
  \label{fig:prism_spectra}
\end{figure}

When instructed to do so in the configuration file, the eMPT gathers the ensemble of spare slitlets, checks their (nodded) slitlets for the presence of targets listed in the input catalog, and passes the set of empty slitlets to the matrix algorithm to determine the optimal subset of sky spectra that can still fit on the detector without overlapping with the target spectra already in place. Figure~\ref{fig:prism_spectra} is an extreme example of this process at work.

{The eMPT automates this process in an optimal way, whereas in the MPT, the user must manually identify and select, one by one, any empty MSA shutters in the MSA Configuration Editor to use in the calculation of a master background sky spectrum for background subtraction. }

\subsection{Scripting the work flow}
\label{sec:script}

The compiled eMPT modules are standard \textit{gfortran} executables that read in the name of the relevant controlling configuration file from the command line. Once the user is comfortable with the eMPT workflow and the various parameter choices that steer it, the entire eMPT run can easily be captured in a simple shell script that executes the modules in the correct sequence (optionally interlaced with the user's own software components).
Since such a scripted eMPT run typically completes in a matter of minutes, this is a good way to prepare and quickly process a given MSA observation once the roll angle at which it will be executed is assigned by STScI (Sect.~\ref{sec:roll}). 

A simple Python wrapper of the Fortran modules is available for running the eMPT in batch mode, designed primarily for experienced users of the software who require many repeated, exploratory trials with different starting parameters set for each. 

\section{Conforming to the Astronomer's Proposal Tool }

The eMPT software suite and the algorithms it employs are designed to produce output in the form of MSA pointings and associated MSA masks that can readily be imported into the STScI APT. This necessitates that the eMPT conform with a number of conventions employed in the APT system. 

\subsection{The roll orientation of the MSA}
\label{sec:roll}

One such aspect of the STScI APT that is inconvenient for an instrument like NIRSpec with its unprecedented complexity, is the manner in which the APT handles the roll angle specifying how the MSA is orientated on the sky.

The precise placement of the MSA on the sky is specified by the celestial coordinates ($\alpha_p,\delta_p$) at which the so-called MSA reference point (nominally the geometrical center of the MSA field of view) is pointed, together with a specification of how the MSA is rotated around this reference position -- usually quantified by the position angle, $\phi_p$, of the MSA y-axis projected to the sky, measured at the MSA reference point. 

NIRSpec MSA observations are, for obvious reasons, very sensitive to rotation. The field of view spanned by the MSA measures 3\farcm6$\times$3\farcm4 on the sky, and the open area of its shutters projects to 199~mas\,$\times$\,463~mas. It follows that a rotation of the MSA around its center by a mere $\Delta\phi_p\simeq$28\arcsec  is sufficient to shift a target located within a shutter situated in the far corner of the field of view by $\simeq$20~mas, or one tenth of the slit width. Reflecting this sensitivity, the NIRSpec target acquisition approach developed for MSA observations  \citep{keye18} performs an active roll correction, in addition to the more customary ``pitch and yaw'' correction of the telescope pointing, to ensure that the targets to be observed are placed within their intended respective open shutters to the required accuracy.

{At the same time, the need to keep the telescope in a perpetual shadow behind its sunshade makes the JWST observatory itself highly roll-constrained -- to the point where JWST users are allowed to specify only in exceptional scientifically justified circumstances the preferred roll angle of their observations, since doing so effectively schedules when the observation can occur.} The majority of NIRSpec MSA users are therefore only assigned a final roll angle by STScI once their observations have been placed on the observatory timeline.

\begin{figure}
  \resizebox{\hsize}{!}{\includegraphics{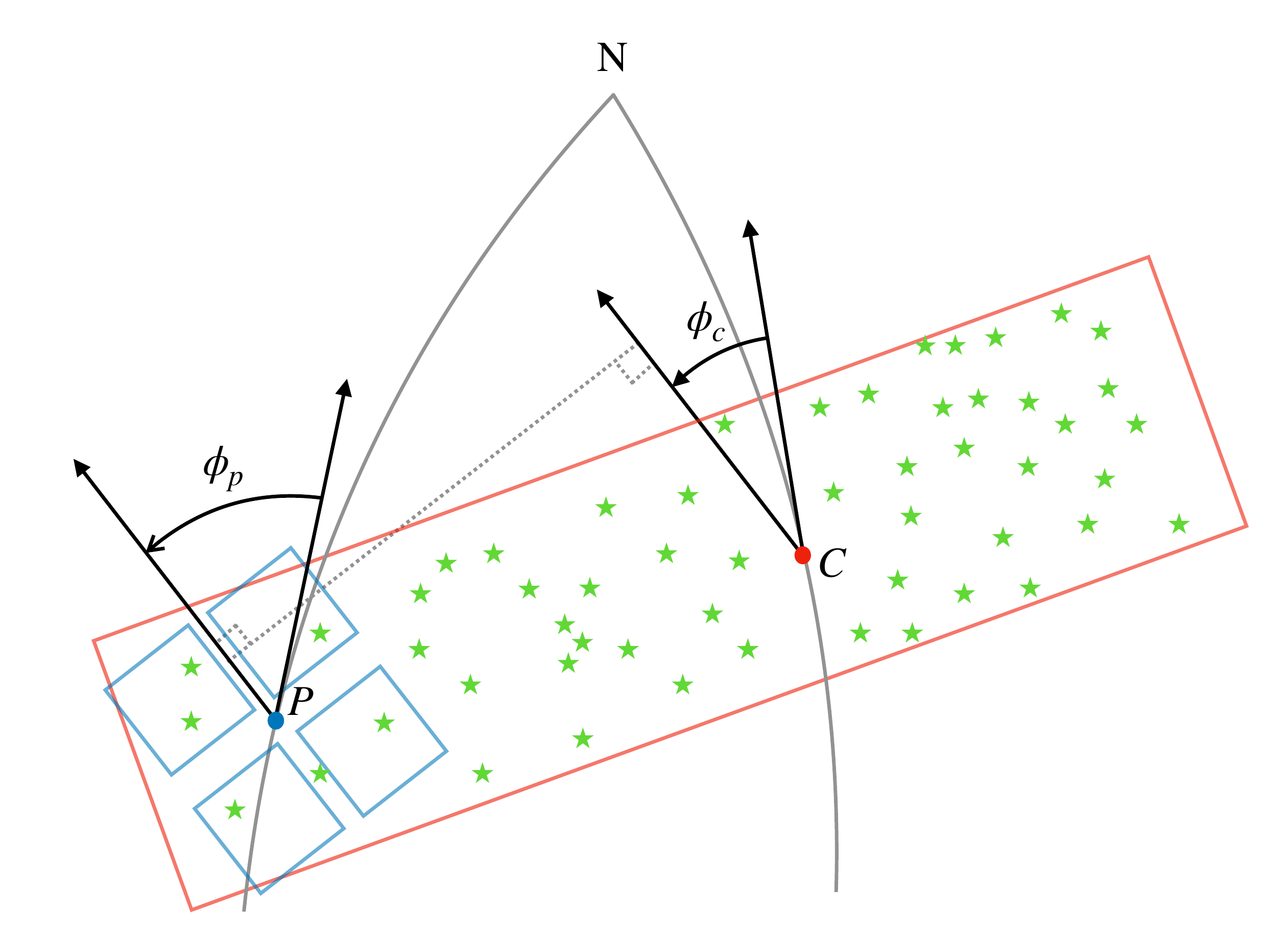}}
  \caption{Schematic illustration of the difference in MSA roll angle measured at the MSA reference point, $P$, and at the input catalog median reference position, $C$.}
  \label{fig:e}
\end{figure}

However, when STScI allocates a roll angle to a given MSA observation, it is not the value of the angle $\phi_p$ of interest that is provided, but rather the angle $\phi_c$, the MSA roll angle measured at the so-called catalog reference position, defined by the median location of {all} objects contained in the target input catalog entered into the APT by the user at the proposal stage (see Fig.~\ref{fig:e}).

Fortunately, the difference between the two roll angles $\phi_p$ and $\phi_c$ is predictable, and reflects the change in bearing toward the north, measured along the great circle connecting the two reference points. If $(\alpha_c,\delta_c)$ denotes the position of the catalog median reference position, the relationship between the actual and assigned MSA orientation for the MSA pointing ($\alpha_p, \delta_p$) can be shown to be $\phi_p =  \phi_c + \Delta \phi$, where
\begin{equation}
\Delta\phi =  \arctan({{\sin(\alpha_p\!-\!\alpha_c) (\sin\delta_c+\sin\delta_p})\over {\cos\delta_c  \cos\delta_p+\cos(\alpha_p\!-\!\alpha_c)(1+\sin\delta_c \sin\delta_p)}})
\label{eq:eq_of_e}
.\end{equation}
It follows from this expression that the farther apart the two reference points are from each other, and/or the closer the target field is to the celestial poles, the greater the difference will be between the two roll angles $\phi_c$ and $\phi_p$.

This feature of the STScI APT is not immediately visible to the user in the MPT, as it silently orients the MSA at any given pointing to an actual roll angle $\phi_p$ that differs from the officially allocated angle $\phi_c$ in accord with Eq.~(\ref{eq:eq_of_e}). Although this involuntary forced shift in roll angle is generally small and inconsequential for an imaging instrument, it can be quite significant for NIRSpec. Regrettably, there is no choice but to mirror this behavior in the eMPT software in order to enable its optimized MSA mask designs to be readily imported into the APT.

A practical example serves to illustrate the above described effect:  a user wishes to obtain NIRSpec MSA spectra of a number of remote galaxies in the GOODS-South Legacy Field and loads the full \cite{whit19} catalog into the eMPT or MPT. This catalog has 186\,474 entries and extends over approximately 24\farcm8$\times$23\farcm4 on the sky, and its median reference position is located at ($\alpha_c, \delta_c$)=(53\fdg142334,-27\fdg808222). In the case where the targets of interest happen to be located in the northeast corner of the catalog near, say, ($\alpha_p, \delta_p$)=(53\fdg264126,-27\fdg684844), 9\farcm8 away from the catalog reference position, then the actual MSA roll angle will, per Eq.~(\ref{eq:eq_of_e}), deviate from the official allocated one by $\Delta\phi=-204$\arcsec, which is substantial enough to rotate many targets out of their intended shutters. The implication is that if the user had, instead, first trimmed down the size of the catalog by entering into the MPT a circular 3\arcmin \, radius cut-out sub-catalog covering only the region of interest, then  $\phi_p$ would lie much closer to $\phi_c$ and a significantly different set of targets would be observable for the exact same pointing and assigned roll angle. 

Thus, a counterintuitive situation arises where the question of which targets are observable with the MSA at any given pointing and assigned roll angle, is not solely determined by the availability of the scientifically interesting targets contained within the MSA field of view, but also by {all} the other catalog entries that are not being observed, even including those that are located well outside the field of view of the MSA. Having the detailed MSA performance tied to its full input catalog in such a fashion is undesirable considering that the MSA is only capable of observing up to $\simeq$200 targets simultaneously in any given exposure \citep{ferr22} -- corresponding to fewer than $\simeq$0.1\% of all catalog entries in the example given above. Employing a symmetrical, trimmed-down catalog centered on the pointing of interest may, however, not be an option in cases where several dithered or tiled MSA exposures of the same field are planned. In these cases, it is unavoidable that the individual dithered or tiled exposures all occur at slightly different values of $\phi_p$ for the fixed assigned value of $\phi_c$.

NIRSpec MSA observers also need to be wary of the reverse situation where any change made by the user to the input catalog after its entry into the MPT or eMPT (such as eliminating superfluous targets and/or adding reference stars needed for target acquisition), in principle shifts the catalog reference position ($\alpha_c, \delta_c$), which in turn rotates the MSA per Eq.~(\ref{eq:eq_of_e}), thereby rendering the original optimized slit mask design invalid. Fortunately, the speed of the eMPT and its easily scripted workflow (Sect.~\ref{sec:script}) make this obstacle easy to work around through iteration, especially if the initial catalog is not too irregularly shaped. 

\subsection{Differential velocity aberration}

The impact of the bulk velocity aberration caused by the movement of the JWST spacecraft is an all-sky distortion effect that is handled at the spacecraft level and is therefore invisible to the NIRSpec user. When the MSA is to be pointed at a given position ($\alpha_p, \delta_p$) on the sky, the telescope is oriented such that the MSA reference point is placed at the aberrated location of ($\alpha_p, \delta_p$) at the time of the observation; however, this only corrects for the velocity aberration at the center of the MSA. To place the desired targets as accurately as possible within their ~199 mas-wide shutters necessitates that the {differential} velocity aberration (DVA) over the full 3.6$\times$3.4~arcmin$^2$ field of view of the MSA also be considered. 

The mathematics of the DVA correction are discussed by \cite{cox03} and \cite{hen10}, who show that the DVA manifests itself as a simple uniform magnification or demagnification of the view of the sky seen by the telescope around the pointing corrected for bulk aberration. In the nonrelativistic regime applicable here, the associated magnification factor is simply
\begin{equation}
M_{DVA}\simeq\frac{1}{1 - (v/c)\cos\theta}
\label{eq:dva}
,\end{equation}
where $v/c$ is the barycentric velocity of the JWST spacecraft relative to the speed of light and $\theta$ is the angle between the JWST velocity vector and the direction to the target point being observed.

Since the peculiar velocity of JWST in its slow halo orbit around L2 is comparatively small (<0.5~km~s$^{-1}$), the motion of the JWST spacecraft can, to first order, be thought of as orbiting the Sun at a constant speed of $v\simeq30$~km~s$^{-1}$, with a velocity vector trailing the Sun by $\simeq$90\degr in the ecliptic plane.   

It follows from Eq. (\ref{eq:dva}) that the DVA field magnification is largest ($M_{DVA}\simeq1.0001$) when looking along the velocity vector in the ecliptic plane ($\theta\simeq$0\degr). Conversely, the DVA demagnification is the greatest ($M_{DVA}\simeq0.9999$) when looking opposite the velocity vector ($\theta\simeq$180\degr), and has no effect ($M_{DVA}=1.0000$) when observing toward ether of the ecliptic poles ($\theta\simeq$90\degr). Thus, the, at most, one part in 10\,000 relative DVA correction has a small, but not entirely negligible, effect on NIRSpec -- in the worst case amounting to intra-shutter displacements of $\pm14$~mas at the extreme corners of the 3.6$\times$3.4~arcmin$^2$ MSA field of view.

In the eMPT, the DVA correction is applied by scaling the tangential coordinates of the targets ($x_\alpha$,$y_\delta$) calculated from Eqs. (\ref{eq:tcoordx}) and (\ref{eq:tcoordy}), using the MSA pointing ($\alpha_p,\delta_p$) as the tangent point by the factor $M_{DVA}$ calculated from Eq.~(\ref{eq:dva}), for the value of $\theta$ entered by the user and assuming a constant spacecraft velocity of $v=30.0$~km~s$^{-1}$. The STScI MPT employs a more accurate velocity vector based on the JWST spacecraft ephemerides, but calculates the angle $\theta$ with respect to the input catalog reference position ($\alpha_c,\delta_c$) discussed above, rather than the actual MSA pointing (although the DVA correction is correctly applied using ($\alpha_p,\delta_p$) as the origin). However, since the magnification factor $M_{DVA}$ according to Eq. (\ref{eq:dva}) is a slowly changing function of $\theta$, this inconsistent approach does not have much of a practical impact in most circumstances.

\subsection{Importing eMPT output into the APT/MPT}

The eMPT output that can be directly imported into the APT to generate a customized APT MOS observation includes a STScI-formatted configuration file (\textit{shutters.csv}) that describes the MSA configuration of the optimized pointing(s), that is, which shutters are to be commanded open and which are to remain closed; the catalog of targets observed at the optimized pointing(s) (\textit{observed-targets.cat}); along with relevant pointing summary information that must be entered into corresponding fields of the APT (\textit{pointing-summary.txt}).

General instructions for creating an APT MOS observation from a custom-made MSA configuration without the use of the MPT Planner can be found in the JWST online documentation on the page entitled ``Custom MOS Observations using the MSA Configuration Editor.'' However, as also warned there, this can be a very tedious process and only intended for experienced users. For this reason, after first importing the input catalog into the APT using the MSA Source Catalog importer, it is advised to take the simpler and quicker path of entering all of the relevant eMPT output into the Observations pane, in the Configurations/Pointings section, and then utilizing the ``Review in MPT'' function to switch over to the MPT Planner to replan as required to match the imported configuration. This preferred procedure for importing the eMPT output into the APT is described in full in the eMPT User Guide.

\label{sec:sw}

\section{Summary and conclusions}

The eMPT software suite is offered to JWST observers as a supplement to the STScI MPT for designing the most scientifically optimal MSA configurations for their planned NIRSpec MOS observations.\ It features robust and novel algorithms and provides great flexibility and customizability to the user due to its modular architecture. It utilizes an up-to-date, in-flight version of the NIRSpec instrument model, and meticulously ensures maximal scientific return for a given observing program by carefully prioritizing targets as desired and maximizing the number of spectra that can cleanly and safely be dispersed together onto the detector arrays in a single exposure. 

The eMPT is capable of matching the \(\sim \)250,000 individually addressable micro-shutters of the NIRSpec MSA to the targets listed in a user-supplied prioritized input catalog, producing output that can be entered directly into the STScI APT. Compared to the MPT, the eMPT offers additional user options for dither selection and an automatic utilization of spare, empty MSA shutters for a master background calculation.

In its present modular form, the eMPT software suite consists of seven independent modules that are run in strict sequence, each step of which the user can interactively examine and approve and optionally rerun with modified parameter settings before proceeding to the next step. Alternatively, a full automatic run of the software can be quickly executed from the command line when only the final output summary, diagnostic, and plot files need to be examined. There is also a user-configurable template script available for running the eMPT in batch mode, designed for experienced users who may need to run many exploratory trials in succession with different starting parameters set for each.

While the eMPT was developed to address the specific scientific needs and concerns of the NIRSpec GTO program, it ended up growing and evolving into something larger and more generally applicable than originally envisioned, in the space provided by the repeated delays of the JWST launch in recent years. Therefore, it is being shared with the JWST NIRSpec observer community at large for all to benefit.

\begin{acknowledgements}
The eMPT software suite is the result of years of discussions on how best to optimize the use of the novel Micro-Shutter Array that have taken place within the NIRSpec GTO Team. Special thanks are due to Diane Karakla and Gary Curtis of STScI for their openness and patience in explaining the inner workings of the APT/MPT and participating in many mutually beneficial exchanges, tests and discussions. We are especially indebted to Bernd Husemann for articulating a specialized and important user case of the eMPT that pushed us to pursue his suggestion of modularizing the eMPT, which greatly enhanced the flexibility and utility of the system, and made it much easier to maintain.\\ 
\\
The Cosmic Dawn Center (DAWN) is funded by the Danish National Research Foundation under grant No. 140\\
\\

SA acknowledges AEI grant PID2021-127718-NB-I00
\end{acknowledgements}

\bibliographystyle{aa}
\bibliography{empt_refs} 

\end{document}